\documentclass[a4paper,12pt]{article}

\usepackage{graphics}
\usepackage{latexsym}
\newcommand{\ap}{\alpha^{\prime}}

\newcommand{\cB}{\mathcal{B}}

\newcommand{\cM}{\mathcal{M}}

\newcommand{\cQ}{\mathcal{Q}}

\newcommand{\zetto}{\mathbf{Z}}
\newcommand{\aaru}{\mathbf{R}}

\begin{document}
\begin{titlepage}
\thispagestyle{empty}
\begin{flushright}
UT-944 \\
hep-th/0106068 \\
June, 2001 
\end{flushright}

\vskip 1.5 cm

\begin{center}
\noindent{\textbf{\LARGE{Survey of the Tachyonic Lump \vspace{0.5cm}\\ 
in Bosonic String Field Theory }}}
\vskip 1.5cm
\noindent{\large{Kazuki Ohmori}\footnote{E-mail: 
ohmori@hep-th.phys.s.u-tokyo.ac.jp}}\\ 
\vspace{1cm}
\noindent{\small{\textit{Department of Physics, Faculty of Science, University of 
Tokyo}} \\ \vspace{2mm}
\small{\textit{Hongo 7-3-1, Bunkyo-ku, Tokyo 113-0033, Japan}}}
\end{center}
\vspace{1cm}
\begin{abstract}
We study some properties of the tachyonic lumps in the level truncation scheme 
of bosonic cubic string field theory. We find that several gauges work well and that 
the size of the lump as well as its tension 
is approximately independent of these gauge choices at level (2,4). 
We also examine the fluctuation spectrum around the lump solution, and find that 
a tachyon with $\ap m^2=-0.96$ and some massive scalars 
appear on the lump world-volume. This result 
strongly supports the conjecture that a codimension 1 lump solution is identified with a 
D-brane of one lower dimension within the framework of bosonic cubic string field theory.  
\end{abstract}
\end{titlepage}
\newpage
\baselineskip 6mm


\section{Introduction}
In the past years, the tachyonic lump solutions~\cite{HK,lump,MSZ,Toy}\footnote{Recently, tachyon 
lump solutions on a curved background was considered in~\cite{Michi}.} as well as the tachyon 
potential~\cite{KS,SZ,pot} have been worked out in bosonic cubic string field theory~\cite{Wit} using 
the level truncation scheme.\footnote{For their counterparts in superstring field theory 
see~\cite{supers,KO2}, and these works are reviewed in~\cite{KO1,Ber}.} However, these calculations 
were carried out only in the Feynman-Siegel gauge. Although its validity has 
been checked in~\cite{FS}, it would be interesting to see whether other gauge choices are possible. 
In fact, it has recently been found in~\cite{ET} that the Feynman-Siegel gauge condition has a finite 
range of validity, and that some other gauge choices lead to similar tachyon potentials to that 
in the Feynman-Siegel gauge. It is the problem of choosing the gauge fixing condition(s) that we will 
address in the first half of this paper. We will see that we can successfully construct lump 
solutions with nearly correct tensions in several gauges, and that the widths of the lumps are 
independent of these gauge chioces at least at level (2,4). 

Concerning this point, in the recent studies of so-called `vacuum string field 
theory'\cite{RSZ1,sliver,vsft} it was guessed that the width of the 
lump solution was a gauge-dependent quantity~\cite{sliver}. In more detail, 
let us consider the center-of-mass coordinate $x^{\mu}$ of the string and its conjugate momentum $p^{\mu}$, 
obeying the canonical commutation relation 
\begin{equation}
[\widehat{x}^{\mu},\widehat{p}^{\nu}]=i\eta^{\mu\nu}. \label{eq:A}
\end{equation}
Combining these two operators, we can move to the oscillator representation~\cite{GJ} of the 
zero modes as~\cite{sliver}
\begin{equation}
a_0^{\mu}=\frac{1}{2}\sqrt{b}\widehat{p}^{\mu}-\frac{i}{\sqrt{b}}\widehat{x}^{\mu},\quad 
a_0^{\mu\dagger}=\frac{1}{2}\sqrt{b}\widehat{p}^{\mu}+\frac{i}{\sqrt{b}}\widehat{x}^{\mu}.
\label{eq:B}
\end{equation}
It is easily found that they satisfy the commutation relation $[a_0^{\mu},a_0^{\nu\dagger}]=
\eta^{\mu\nu}$ for the creation-annihilation operators irrespective of the value of the 
newly introduced parameter $b$. Moreover, it has been observed in~\cite{sliver} that the value of 
the ratio of the lump tensions seems to converge to the expected single value for all $b$, while 
the extent of the lump in spacetime clearly depends on $b$. From these facts, the authors 
of~\cite{sliver} has proposed that 
the solutions of different widths are related to each other by the gauge 
transformation which does not affect the ghost part of the string field. Though our result suggests 
that the size of the lump cannot be changed by gauge transformation in the `ordinary' cubic 
string field theory,\footnote{By `ordinary cubic string field theory' we mean the one in which the 
kinetic operator is the usual BRST operator $Q_B$ in the flat spacetime, as opposed to the 
vacuum string field theory.} 
it does not immediately contradict the above proposal because the relation 
between vacuum string field theory and the ordinary cubic string field theory is yet to be clarified. 
Since some of the symmetries in the closed string vacuum are broken in a D-brane background, it is 
conceivable that the ordinary cubic string field theory we will use has fewer gauge degrees of 
freedom than the vacuum string field theory. If the gauge symmetry which actually changes the width 
of the lump is broken in the presence of a D-brane background, we cannot see such an effect 
using the ordinary cubic open string field theory on a D-brane. 
\medskip

In the second half of this paper, we consider the fluctuation spectrum around the codimension-1 
lump solution constructed in the modified level truncation scheme~\cite{MSZ}. 
By now, it has been verified that the spacetime-independent non-perturbative vacuum solution 
in the ordinary cubic string field theory does not support any perturbative 
open string excitations~\cite{noopen} 
within the level truncation scheme. On the other hand, if the conjecture that the tachyonic lump 
solutions represent the lower-dimensional D-branes~\cite{lower} is true, the fluctuation spectrum 
around the lump solution should agree with that of the corresponding D-brane. Analyses of this kind 
have so far been done in the tachyonic $\phi^3$ scalar field theory 
model\footnote{The fluctuation spectra on the lumps (or kinks) were also studied in $p$-adic 
string theory~\cite{padic} and in field theory models for tachyon dynamics~\cite{MZ}.} \cite{HK,Toy}, the 
result being that the spectrum on the lump contains a tachyon with $\ap m^2=-5/4$, a massless 
scalar and an approximately massless gauge field (it is reviewed in section 4.1 of \cite{KO1}). 
At this level of approximation, the mass of the tachyon is not so close to the expected 
value $\ap m^2=-1$ and the tension of the lump is only 78\% of the correct answer. We will 
examine the fluctuation spectrum on the lump solution to a higher degree of accuracy by taking 
into account the scalar component fields at level 2, and find that there is a tachyon with 
mass $\ap m^2\simeq -0.96$.
\medskip

This paper is organized as follows. In section 2 we discuss the dependence of the tachyonic 
lump on the gauge choice at level (2,4) approximation. In section 3, we determine the low-lying 
mass spectrum around the lump solution in the Feynman-Siegel gauge, and find that a tachyon 
appears with the nearly correct mass squared. Section 4 includes a summary of our results and 
discussions on some points. Numerical data for expectation values and quite lengthy expressions 
for the action are collected in Appendices. 

\section{Tachyonic Lumps in Various Gauges}\label{sec:gauge}
In this section we study the dependence of the lump solutions on the gauge choice. 
We consider the possibility of imposing gauge fixing conditions other than 
the Feynman-Siegel gauge $b_0|\Phi\rangle=0$ on the string field as well as that of leaving 
the gauge unfixed. 

In the modified level truncation scheme introduced in~\cite{MSZ}, momentum in the $X$-direction, 
in which we will construct non-trivial field configurations, is discretized as $p=n/R$ with 
$n$ integer by compactifying the $X$-direction on a circle of radius $R$. This compact momentum 
is taken to contribute to the level as 
\begin{equation}
\mathrm{level}=\frac{\ap}{R^2}n^2+h_N+1, \label{eq:C}
\end{equation}
where $h_N$ denotes the conformal weight of the state excluding the contribution from the 
momentum factor. (For more details, see~\cite{MSZ} and chapter 4 of~\cite{KO1}.) To this end, 
we consider the level (2,4) truncation approximation. At this level, the string field 
without gauge fixing, restricted to the twist-even sector, is expanded as 
\begin{equation}
|\Phi\rangle=\sum_n\phi_n\ c_1|n\rangle+u\ c_{-1}|0\rangle+v\ L_{-2}^Xc_1|0\rangle+w\ L_{-2}^{\cM}
c_1|0\rangle+r\ b_{-2}c_0c_1|0\rangle, \label{eq:D}
\end{equation}
where $|n\rangle=e^{i\frac{n}{R}X(0)}|0\rangle$ and $|0\rangle$ is the $SL(2,\aaru)$-invariant 
vacuum. We have denoted by $\cM$ the 25-dimensional manifold excluding $X$, and by $L_m^{\cM}$ the 
Virasoro generators of the conformal field theory of central charge 25 associated with $\cM$. 
In finding the lump solution which is symmetric under the reflection $X\to -X$, we can put 
the following constraints on $\phi_n$'s, 
\begin{eqnarray}
\phi_0&=&\tau_0, \label{eq:E} \\
\phi_n&=&\phi_{-n}=\frac{1}{2}\tau_n \qquad \mathrm{for} \quad n\ge 1, \nonumber
\end{eqnarray}
so that $\phi_n|n\rangle+\phi_{-n}|-n\rangle=\tau_n\cos\frac{n}{R}X(0)|0\rangle$. The range of $n$ over 
which the summation in eq.(\ref{eq:D}) is taken depends on the value of $R$. In this section our 
choices of $R$ are $\sqrt{3\ap}$ and $2\sqrt{2\ap}$, in which cases 
\begin{eqnarray}
-2\le n\le 2 \quad &\mathrm{for}& \ R=\sqrt{3\ap}, \nonumber \\
-4\le n\le 4 \quad &\mathrm{for}& \ R=2\sqrt{2\ap}, \label{eq:F}
\end{eqnarray}
at level 2. 
Due to the reality condition $(\mathrm{bpz}|\Phi\rangle)^{\dagger}=|\Phi\rangle$ on the string 
field, the component fields $\tau_n,u,v,w,r$ must be real. 

Substituting the level-expanded string field (\ref{eq:D}) into the cubic action 
\begin{equation}
S=-\frac{1}{g_o^2}\left(\frac{1}{2}\langle\Phi|Q_B|\Phi\rangle+\frac{1}{3}\langle
\Phi,\Phi,\Phi\rangle\right) \label{eq:G}
\end{equation}
on a D$p$-brane, we have found the level (2,4) truncated gauge-unfixed action to be 
\begin{eqnarray}
\frac{-g_o^2S}{2\pi RV_p}&=&-\frac{1}{2}\tau_0^2+\sum_{n\ge 1}\frac{1}{4}\left(\frac{\ap}{R^2}
n^2-1\right)\tau_n^2-\frac{1}{2}u^2+\frac{1}{4}v^2+\frac{c^{\cM}}{4}w^2+2r^2-3ur \nonumber \\
& &{}+\frac{1}{2}vr+\frac{c^{\cM}}{2}wr+\frac{1}{3}\sum\delta_{n_1+n_2+n_3}K^{3-\frac{\ap}{R^2}
(n_1^2+n_2^2+n_3^2)}\phi_{n_1}\phi_{n_2}\phi_{n_3} \nonumber \\ & &+K^3\Biggl[\frac{11}{27}\tau_0^2 u
+\frac{11}{54}\sum_{n\ge 1}K^{-\frac{2\ap}{R^2}n^2}\tau_n^2u \nonumber \\ & &\qquad -\frac{5}{54}\tau_0^2v
+\frac{1}{2}\sum_{n\ge 1}\left(\frac{16\ap}{27R^2}n^2-\frac{5}{54}\right)K^{-\frac{2\ap}{R^2}n^2}
\tau_n^2v \nonumber \\ & &\qquad -\frac{5}{54}c^{\cM}\ \tau_0^2 w-\frac{5}{108}c^{\cM}\sum_{n\ge 1}
K^{-\frac{2\ap}{R^2}n^2}\tau_n^2w \nonumber \\ & &\qquad -\frac{16}{27}\tau_0^2r-\frac{8}{27}\sum_{n\ge 1}
K^{-\frac{2\ap}{R^2}n^2}\tau_n^2r \nonumber \\ & &\qquad +\frac{19}{243}\tau_0u^2+\frac{179}{972}
\tau_0v^2+\frac{c^{\cM}}{729}\left(\frac{25}{4}c^{\cM}+128\right)\tau_0w^2+\frac{64}{243}\tau_0r^2 
\nonumber \\ & &\qquad -\frac{55}{729}\tau_0uv-\frac{55}{729}c^{\cM}\ \tau_0uw+\frac{32}{81}\tau_0ur 
\nonumber \\ & &\qquad +\frac{25}{1458}c^{\cM}\ \tau_0vw+\frac{80}{729}
\tau_0vr+\frac{80}{729}c^{\cM}\ \tau_0wr\Biggr], \label{eq:H}
\end{eqnarray}
where $c^{\cM}=25$, $K=3\sqrt{3}/4$, and note that our normalization convention is such that 
every component field is dimensionless and that we are not setting $\ap$ to any fixed value. 
In spite of leaving the gauge degrees of freedom unfixed, the procedure of level truncation 
actually breaks the gauge invariance, so that the equations of motion obtained by varying the 
action~(\ref{eq:H}) have a discrete set of solutions, instead of a continuous family. 
Hence we can find the \textit{isolated} `closed string vacuum' solution and the lump solutions for two 
values~(\ref{eq:F}) of radius by solving the simultaneous equations 
of motion numerically, even without any gauge fixing. 
The expectation values found this way are given in Table~\ref{tab:TAa} in 
Appendix~\ref{sec:appA}. Putting these values back to the above action, we can obtain the 
numerical values for the depth of the tachyon potential and for the tensions of the lump 
solutions using the following formulae~\cite{MSZ,KO1}, respectively, 
\begin{eqnarray}
f(\Phi_{\mathrm{vac}})&\equiv&-\frac{S(\Phi_{\mathrm{vac}})}{2\pi RV_p\tau_p}=2\pi^2\times
\mbox{(r.h.s. of eq.(\ref{eq:H}))}, \label{eq:I} \\
r(R,\Phi_{\mathrm{lump}})&=&\frac{R}{\sqrt{\ap}}\Bigl(f(\Phi_{\mathrm{lump}})-f(
\Phi_{\mathrm{vac}})\Bigr), \label{eq:J}
\end{eqnarray}
where we have denoted by $\tau_p$ the tension of the original D$p$-brane, by $\Phi_{\mathrm{vac}}$ 
the closed string vacuum configuration of the string field, and by $\Phi_{\mathrm{lump}}$ 
the lump configuration. We have used the relation~\cite{univ} 
\[ \tau_p=\frac{1}{2\pi^2g_o^2} \]
between the D$p$-brane tension and the open string coupling on the D$p$-brane in the most 
right hand side of eq.(\ref{eq:I}). The conjectured values for $-f(\Phi_{\mathrm{vac}})$ and 
$r(^{\forall}\! R,\Phi_{\mathrm{lump}})$ are 1. Our results obtained with the gauge left unfixed 
are 
\begin{equation}
f(\Phi_{\mathrm{vac}})=-0.885220, \quad r(\sqrt{3\ap},\Phi_{\mathrm{lump}})=0.869597, \quad 
r(\sqrt{8\ap},\Phi_{\mathrm{lump}})=0.679800. \label{eq:K}
\end{equation}
These values are summarized in Table~\ref{tab:Ta}, together with the gauge-fixed ones 
to be explained below. 
\medskip

Next we consider fixing the gauge. As we mentioned above, the gauge invariance has actually 
been broken by the level truncation. However, this breakdown is not under control so that 
the gauge-fixing is needed to obtain more definite results. The first gauge fixing condition we try 
to choose is the Feynman-Siegel gauge 
\begin{equation}
b_0|\Phi\rangle=0, \label{eq:L}
\end{equation}
whose validity at the linearized level was established in~\cite{SZ}, and its range of validity 
in the non-perturbative case has been argued in~\cite{ET}. In terms of the component fields 
appearing in eq.(\ref{eq:D}), this condition is restated simply as 
\begin{equation}
r=0. \label{eq:T}
\end{equation}
As alternatives we consider the following three choices 
\begin{eqnarray}
b_1|\Phi\rangle&=&0, \nonumber \\
L_2^X|\Phi\rangle&=&0, \label{eq:M} \\
L_2^{\cM}|\Phi\rangle&=&0, \nonumber
\end{eqnarray}
whose component forms are, respectively, 
\begin{eqnarray}
u&=&0, \nonumber \\
v&=&0, \label{eq:N} \\
w&=&0, \nonumber
\end{eqnarray}
as suggested in~\cite{ET}. The validity of these gauges at the linearized level can 
be verified in a similar way to the case of the Feynman-Siegel gauge, at least in the level (2,4) 
truncation scheme. To give an example, let us look at the $b_1$-gauge closely. Suppose 
that the initial string field $|\Phi\rangle$ is not annihilated by $b_1$. Acting on $|\Phi\rangle$ 
with the linearized gauge transformation, we reach the transformed field 
\begin{equation}
|\widetilde{\Phi}\rangle=|\Phi\rangle-Q_B|\Lambda\rangle, \label{eq:O}
\end{equation}
where the gauge parameter $|\Lambda\rangle$ must have ghost number 0. If $|\Lambda\rangle$ is 
taken to satisfy 
\begin{equation}
b_1|\Lambda\rangle=0 \quad \mathrm{and} \quad L_1^{\mathrm{tot}}|\Lambda\rangle=b_1|\Phi\rangle, 
\label{eq:P}
\end{equation}
then it follows that 
\begin{eqnarray*}
b_1|\widetilde{\Phi}\rangle&=&b_1|\Phi\rangle-b_1Q_B|\Lambda\rangle=b_1|\Phi\rangle-
\{b_1,Q_B\}|\Lambda\rangle \\ &=&b_1|\Phi\rangle-L_1^{\mathrm{tot}}|\Lambda\rangle=0,
\end{eqnarray*}
so we have found that there is a representative obeying the gauge fixing condition $b_1|\widetilde{\Phi}
\rangle=0$ in each gauge equivalent class, \textit{if there exists a suitable gauge parameter 
$|\Lambda\rangle$ satisfying~(\ref{eq:P})}. The string field $|\Lambda\rangle$ of ghost number 0 
is expanded as 
\[ |\Lambda\rangle=\sum_n\lambda_n\ b_{-1}c_1|n\rangle+\sum_n\mu_n\ b_{-2}c_1|n\rangle
+\sum_{n\neq 0}\nu_n\ L_{-1}^Xb_{-1}c_1|n\rangle+\cdots . \]
Since we are focusing on the even level fields thanks to the twist symmetry, the first term 
(at oscillator level 1) can be dropped. In addition, the third and the still higher terms do 
not contribute at level 2. Therefore, we have only to consider the 1-parameter gauge degree 
of freedom 
\begin{equation}
|\Lambda\rangle=\mu\ b_{-2}c_1|0\rangle \label{eq:Q}
\end{equation}
at this level of approximation. It is easily found that the above $|\Lambda\rangle$ satisfies 
both conditions in~(\ref{eq:P}) if we take $\mu$ to be equal to $u/3$. Furthermore, it is obvious 
that there is no residual gauge degree of freedom after fixing the value of $\mu$. 
Similarly, we can reach the $L_2^X$- and $L_2^{\cM}$-gauge by taking $\mu=v$ and $\mu=w$, 
respectively, with no residual gauge degrees of freedom left unfixed. 
Putting aside the problem on the validity of these gauge fixing conditions at higher levels, 
we will try to construct lump solutions as well as the tachyon 
potential using these gauge choices. And we will test one more gauge fixing condition
\begin{equation}
B|\Phi\rangle\equiv \left(b_0+\frac{1}{2}(b_2+b_{-2})\right)|\Phi\rangle=0 \label{eq:R}
\end{equation}
which was proposed in~\cite{RSZ1} as a modified version of the Feynman-Siegel gauge. This 
modification is necessary in vacuum string field theory with a non-standard BRST operator 
$\cQ$ made purely out of ghost operators because of the problems described in~\cite{RSZ1}.
Though the ordinary cubic string field theory has no problem about the Feynman-Siegel gauge, 
it is worth trying it in our context. Written in terms of 
the component fields, $B|\Phi\rangle=0$ gives 
\begin{equation}
r=\frac{1}{2}\tau_0. \label{eq:S}
\end{equation}
\smallskip

After putting the gauge fixing conditions (\ref{eq:T}),(\ref{eq:N}) or (\ref{eq:S}) on the 
action~(\ref{eq:H}), we look for solutions by extremizing the resulting action with respect to 
the remaining field variables. The field values which solve the equations of motion are shown 
in Table~\ref{tab:TAa}--\ref{tab:TAc} in Appendix~\ref{sec:appA}, and the values of the lump tensions 
are displayed in Table~\ref{tab:Ta}.
\begin{table}[tb]
	\begin{center}
	\begin{tabular}{|c|c|c|c|}
	\hline
	gauge & depth of & lump tension & lump tension \\ 
	 & the potential & for $R=\sqrt{3\ap}$ & for $R=\sqrt{8\ap}$ \\ \hline \hline
	left unfixed & $-0.885220$ & 0.869597 & 0.679800 \\ \hline
	Feynman-Siegel & $-0.948553$ & $1.02368\phantom{0}$ & $1.04513\phantom{0}$ \\ \hline
	$b_1|\Phi\rangle=0$ & $-0.894765$ & 0.985080 & $1.04153\phantom{0}$ \\ \hline
	$L_2^X|\Phi\rangle=0$ & $-0.926410$ & 0.958814 & 0.959953 \\ \hline
	$L_2^{\cM}|\Phi\rangle=0$ & $-0.900681$ & 0.994212 & $1.04559\phantom{0}$ \\ \hline
	$B|\Phi\rangle=0$ & $-0.935830$ & $1.21536\phantom{0}$ & $1.51654\phantom{0}$ \\ \hline
	\end{tabular}
	\end{center}
	\caption{The numerical values of the depth $f(\Phi_{\mathrm{vac}})$ of the tachyon potential 
	and those of the lump tensions $r(R,\Phi_{\mathrm{lump}})$ for various gauge choices, at 
	level (2,4) approximation.}
	\label{tab:Ta}
\end{table}
From this table, it seems that the middle four gauges are working pretty well, whereas 
there is something wrong with the $B$-gauge and the non-gauge-fixed one in constructing lumps. 
\medskip

\begin{figure}[htbp]
	\begin{center}
	\includegraphics{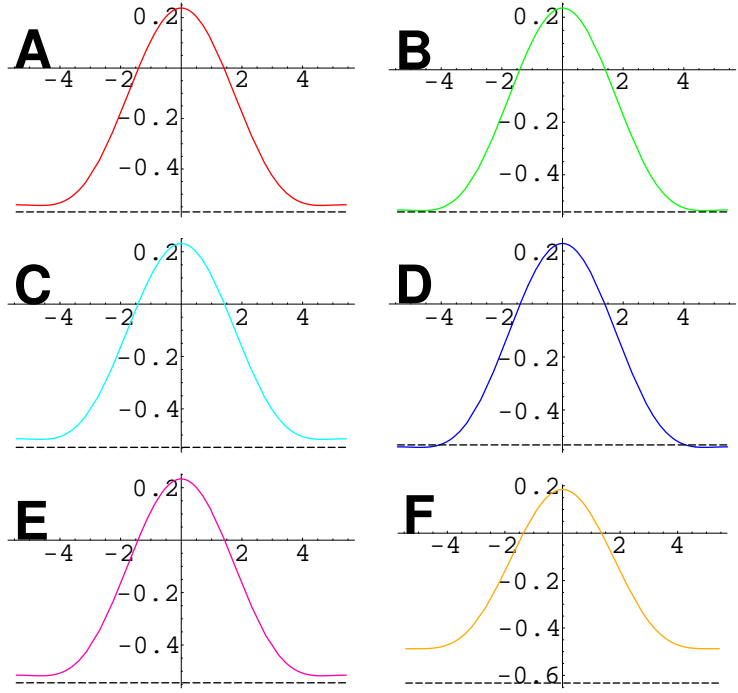}
	\end{center}
	\caption{The tachyon profiles $-t(x)$ for $R=\sqrt{3\ap}$. A: unfixed, 
	B: Feynman-Siegel gauge, C: $u=0$, D: $v=0$, 
	E: $w=0$, F: $B=0$. The dashed lines represent the expectation values of tachyon field at the 
	closed string vacuum.}
	\label{fig:FA}
\end{figure}

\begin{figure}[htbp]
	\begin{center}
	\includegraphics{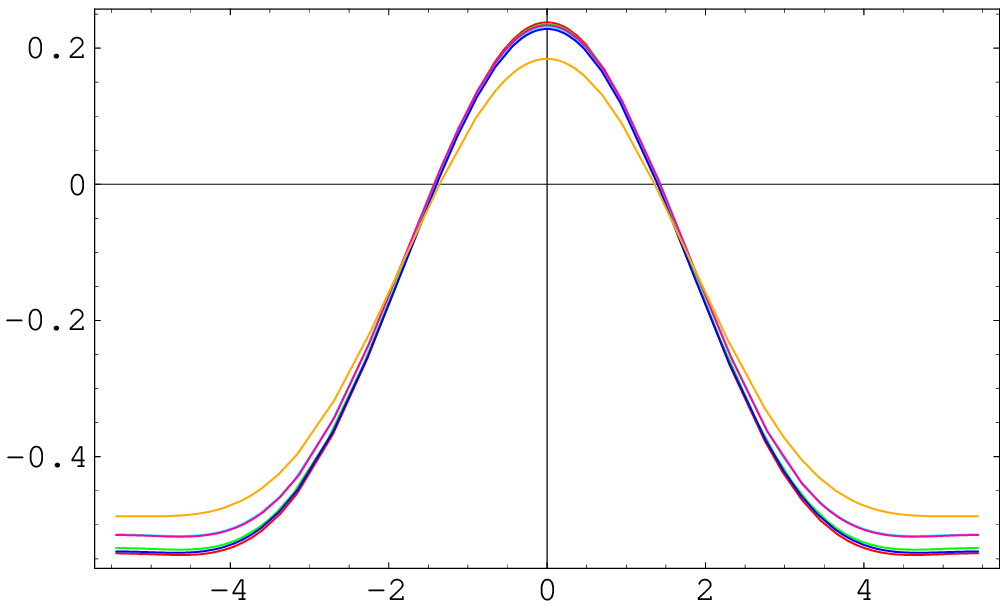}
	\end{center}
	\caption{The tachyon profiles $-t(x)$ for $R=\sqrt{3\ap}$ in various gauges.}
	\label{fig:FC}
\end{figure}

\begin{figure}[htbp]
	\begin{center}
	\includegraphics{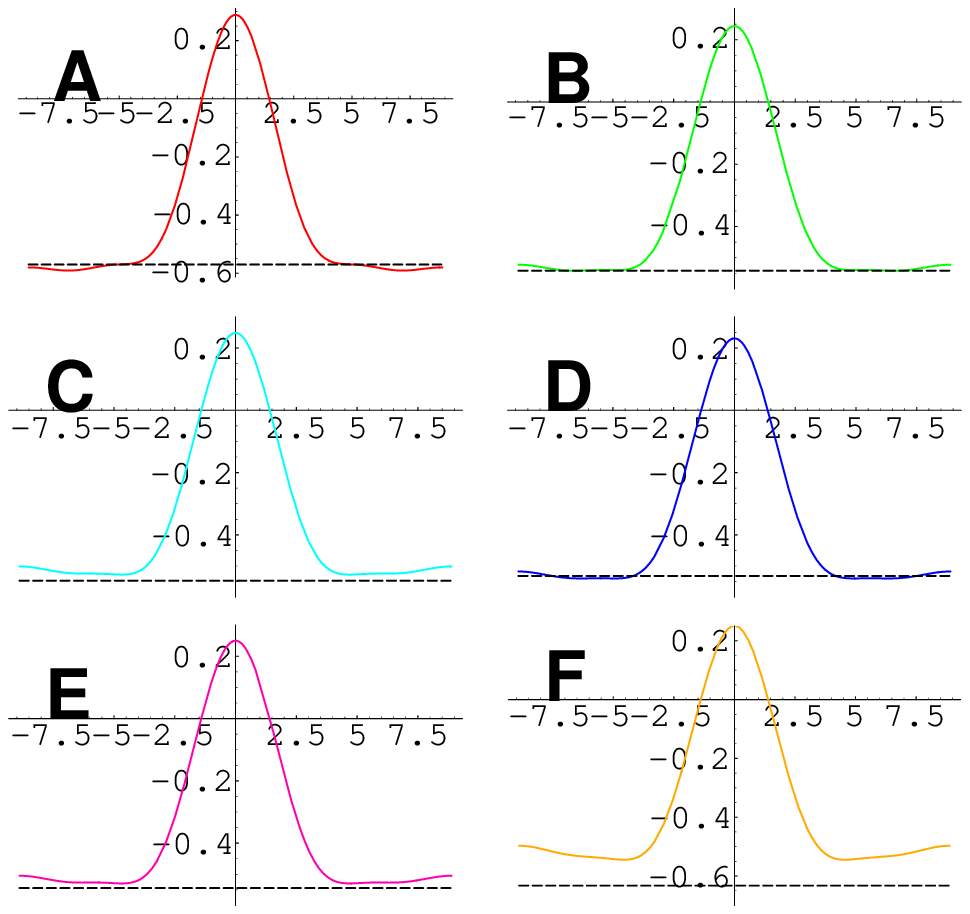}
	\end{center}
	\caption{The tachyon profiles $-t(x)$ for $R=\sqrt{8\ap}$. A: unfixed, 
	B: Feynman-Siegel gauge, C: $u=0$, D: $v=0$, 
	E: $w=0$, F: $B=0$. The dashed lines represent the expectation values of tachyon field at the 
	closed string vacuum.}
	\label{fig:FB}
\end{figure}

\begin{figure}[htbp]
	\begin{center}
	\includegraphics{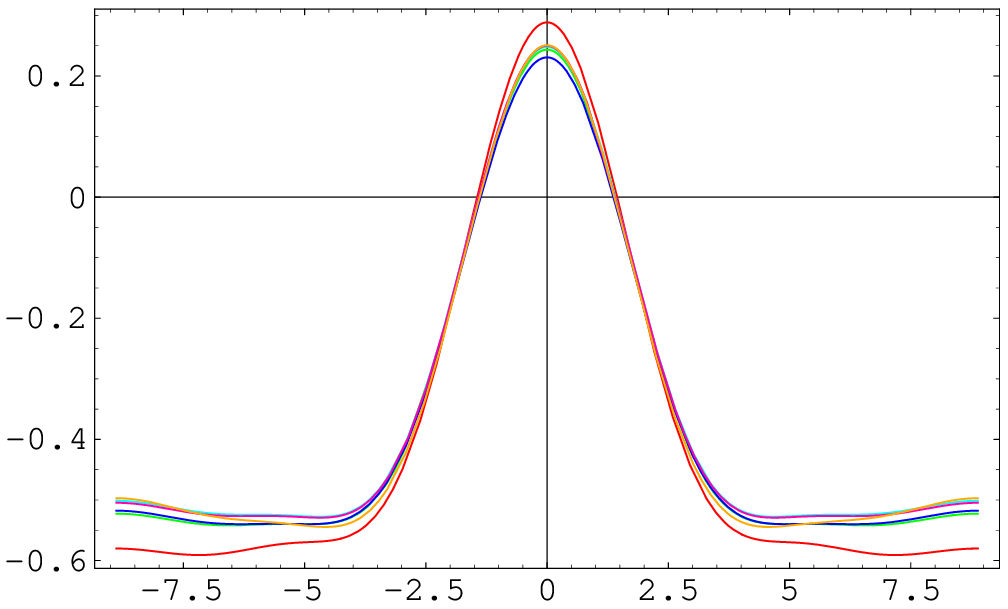}
	\end{center}
	\caption{The tachyon profiles $-t(x)$ for $R=\sqrt{8\ap}$ in various gauges.}
	\label{fig:FD}
\end{figure}

\begin{figure}[htbp]
	\begin{center}
	\includegraphics{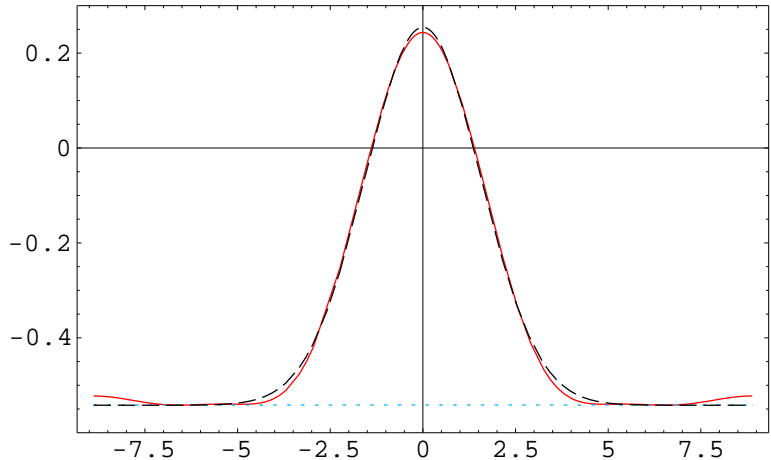}
	\end{center}
	\caption{The result of the fitting of the lump profile $t(x)$ (solid line) 
	with the gaussian $G(x)$ (dashed line): 
	$R=2\sqrt{2\ap}$, Feynman-Siegel gauge.}
	\label{fig:fit}
\end{figure}

Given the fact that the reasonable values for the tension of the lump are reproduced in four 
cases, we wish to compare the profiles of the tachyon field given by 
\begin{equation}
t(x)=\sum_{n\in\zetto}\phi_ne^{i\frac{n}{R}x}=\sum_{n\in\zetto_{\ge 0}}\tau_n\cos\frac{n}{R}x. 
\label{eq:U}
\end{equation}
Substituting the expectation values shown in Table~\ref{tab:TAa}--\ref{tab:TAc} into eq.(\ref{eq:U}), 
we can explicitly find the tachyon profiles in various gauges. Figure~\ref{fig:FA} and 
Figure~\ref{fig:FB} show them put side by side, for $R=\sqrt{3\ap}$ and $R=\sqrt{8\ap}$ 
respectively. And, these profiles are superposed in Figure~\ref{fig:FC} and in Figure~\ref{fig:FD} for 
each radius. 
These figures clearly indicate that the profiles found in different gauges are almost 
identical. To check this quantitatively, we fit each profile~(\ref{eq:U}) with a Gaussian 
curve of the form 
\begin{equation}
G(x)=a+b e^{-\frac{x^2}{2\sigma^2}}. \label{eq:V}
\end{equation}
We have found the resulting values of $\sigma$ to be 
\begin{equation}
\left.
	\begin{array}{|c||c|c|c|c|c|c|}
	\hline
	\mbox{Gauge} & \mbox{left unfixed} & \mbox{Feynman-Siegel} & b_1 & L_2^X & L_2^{\cM} 
	& B \\ \hline 
	R=\sqrt{3\ap} & 1.61433 & 1.61608 & 1.61711 & 1.62301 & 1.61531 & 1.66272 \\ \hline
	R=2\sqrt{2\ap} & 1.56733 & 1.55317 & 1.53030 & 1.55641 & 1.53214 & 1.50111 \\ \hline
	\end{array}
\right. \label{eq:W}
\end{equation}
The full set of values for $(a,b,\sigma)$ is shown in Table~\ref{tab:TAd} in Appendix~\ref{sec:appA}. 
As an illustration, we show in Figure~\ref{fig:fit} the result of the fitting for the solution 
obtained in the Feynman-Siegel gauge for $R=2\sqrt{2\ap}$. 
In order to estimate errors, we quote the following result from~\cite{MSZ}:
\begin{center}
$\sigma=1.52341$ at level (3,6) for $R=\sqrt{3\ap}$ in Feynman-Siegel gauge.
\end{center}
Hence the error originating from the level truncation approximation is estimated to be 
\[ \frac{|1.61608-1.52341|}{1.61608}\simeq 0.057. \]
Since the range of the values (\ref{eq:W}) obtained in various gauges for each fixed radius is well 
within 6\%, our result suggests that the values of $\sigma$ for lump solutions 
found in different gauges agree with each other. 
It has already been pointed out in~\cite{MSZ} that the size $\sigma$ of the lump solution in 
Feynman-Siegel gauge is independent of the radius $R$.\footnote{In table (\ref{eq:W}), one may think 
that there is a significant difference between $\sigma\simeq 1.62$ $(R=\sqrt{3\ap})$ and $\sigma\simeq 
1.55$ $(R=2\sqrt{2\ap})$. We expect this discrepancy to decrease as we increase the truncation level.} 
Besides, we have found that the size of the lump is also independent of the gauge choices, at least 
within the range of our approximation. 

\section{Fluctuation Spectrum Around the Tachyonic Lump}
In this section, we analyze the fluctuation spectrum on the lump world-volume in the 
Feynman-Siegel gauge for $R=\sqrt{3\ap}$: While we expect that any other gauge will do, it seems 
that the Feynman-Siegel gauge has a better convergence property than others in the sense of 
level truncation~\cite{ET}. On one hand too small a value of radius makes the structure of the 
lump vague, and on the other hand too large a value of radius makes the calculations less accurate. 
In general, if we have a solution $\Phi_0$ to the equation of motion available, the cubic action 
for the fluctuation field $\Phi^{\prime}$ expanded around the solution $\Phi_0$ becomes\footnote{The 
form of the action for the fluctuation fields around a solution in Berkovits' superstring field 
theory has recently been discussed in~\cite{Kluson}.}
\[ S(\Phi_0+\Phi^{\prime})=S(\Phi_0)-\frac{1}{g_o^2}\left(\frac{1}{2}\langle \Phi^{\prime}|Q|
\Phi^{\prime}\rangle+\frac{1}{3}\langle\Phi^{\prime},\Phi^{\prime},\Phi^{\prime}\rangle\right), \]
where the new kinetic operator $Q$ is defined by 
\[ Q\Phi^{\prime}=Q_B\Phi^{\prime}+\Phi_0*\Phi^{\prime}+\Phi^{\prime}*\Phi_0, \]
and we can show that $Q$ is also nilpotent, $Q^2=0$. Hence the physical perturbative spectrum 
around the solution $\Phi_0$ is determined by the cohomology of $Q$. For the tachyon vacuum 
solution $\Phi_0=\Phi_{\mathrm{vac}}$, it has numerically been verified that $Q$ has vanishing 
cohomology~\cite{noopen} and, more strongly, it has recently been proposed that, after a suitable 
field redefinition, $Q$ can be brought to a simple form made purely out of ghosts~\cite{RSZ1}. 
On the other hand, for the codimension-1 lump solution $\Phi_0=\Phi_{\mathrm{lump}}$, we expect 
that $Q$ should be again the BRST operator such that the cohomology of $Q$ reproduces the perturbative 
open string spectrum on a D-brane of one lower dimension. Since, however, we have not gotten 
a closed form expression for $\Phi_{\mathrm{lump}}$, we will proceed with the help of the level 
truncation approximation. 
In principle, all we have to do is to rewrite the string field theory action~(\ref{eq:G}) in terms of 
the component fields having the general momentum-dependence, to expand them about the expectation 
values for the lump solution, and to look for zeroes of the quadratic form for the fluctuation 
fields. However, the existence of the off-diagonal pieces arising from the cubic interaction terms 
complicates the analysis, as explained below. 
\medskip

To begin with, let us state our settings. The original D-brane is a space-filling D25-brane 
in the 26-dimensional flat spacetime, and the codimension 1 lump is, of course, to be identified 
with a flat D24-brane. We will focus on the scalar fields up to level 2, restricting to the 
twist-even sector. While it is interesting to incorporate also the twist-odd scalar fields, 
we will not do so because they do not mix\footnote{This in particular 
means that the twist-odd scalars do not contribute 
to the `tachyon' field which will appear on the unstable lump.} with twist-even fields in the 
quadratic terms and, practically, adding these terms makes the calculations much more lengthy. 
Therefore, the expansion of the string field we will consider becomes 
\begin{eqnarray}
|\Phi\rangle&=&\int d^{25}\! k\Biggl(\sum_{n=-2}^2\phi_n(k)\ c_1|n,k\rangle+u(k)\ c_{-1}
|0,k\rangle\label{eq:AC} \\ & &{}+B_{MN}(k)\ \alpha_{-1}^M\alpha_{-1}^Nc_1|0,k\rangle+i
\check{B}_M(k)\ \alpha_{-2}^Mc_1|0,k\rangle\Biggr), \nonumber
\end{eqnarray}
where $k$ is the 25-dimensional momentum vector along $\cM$, $|n,k\rangle=e^{i\frac{n}{R}X(0)+
ik_{\mu}X^{\mu}(0)}|0\rangle$, and $M,N$ run from 0 to 25, while $\mu,\nu$ from 0 to 24 
$(X\equiv X^{25})$. For simplicity, we have assumed that $\cM$ is non-compact flat $\aaru^{1,24}$ 
ignoring the problem that the D-brane has an infinite mass, which would be resolved by compactifying 
all the space directions on a torus of large radii. The reality conditions 
\begin{equation}
\phi_n(k)^*=\phi_{-n}(-k),\quad u(k)^*=u(-k), \quad B_{MN}(k)^*=B_{MN}(-k),\quad 
\check{B}_M(k)^*=\check{B}_M(-k) \label{eq:AD}
\end{equation}
follow from the reality condition imposed on the string field. In this representation, the expectation 
values of the fields corresponding to the lump solution take the forms\footnote{Our metric convention 
is $\eta_{\mu\nu}=\mathrm{diag}(-++\ldots+)$.}
\begin{eqnarray}
\overline{\phi_0(k)}&=&\overline{\tau_0}\delta^{25}(k), \nonumber \\
\overline{\phi_{\pm n}(k)}&=&\overline{\phi_{\pm n}}\delta^{25}(k)=\frac{1}{2}\overline{\tau_n}
\delta^{25}(k) \qquad \mathrm{for}\ \ n=1,2, \nonumber \\
\overline{u(k)}&=&\overline{u}\delta^{25}(k), \label{eq:AE} \\
\overline{\check{B}_M(k)}&=&0, \nonumber \\
\overline{B_{MN}(k)}&=&\left(
	\begin{array}{c|c}
	\frac{\overline{w}}{2}\eta^{(25)}_{\mu\nu} & 0 \\ \hline 
	0 & \frac{\overline{v}}{2}
	\end{array}
\right) \delta^{25}(k), \nonumber 
\end{eqnarray}
where $(\overline{\tau_n},\overline{u},\overline{v},\overline{w})$ are given in Table~\ref{tab:TAa} 
for the Feynman-Siegel gauge, $R=\sqrt{3\ap}$. The reason why we are retaining the vector and the 
tensor fields in eq.(\ref{eq:AC}) is that the longitudinal components of them and the trace of 
$B_{\mu\nu}$, as well as the transverse ($X$-)components, 
behave as scalars and mix with $\phi_n$ and $u$. 
We almost follow the conventions of~\cite{KS}, 
but with a slight difference encountered later. 
\medskip

Substituting the string field~(\ref{eq:AC}) into the cubic action~(\ref{eq:G}) again, we have 
obtained the level (2,4)-truncated action written in terms of the component fields. 
We write down the explicit expression of it in Appendix~\ref{sec:appB}: we have derived it 
using the technology of conservation laws developed in~\cite{RZ}. Given this expression, 
we can obtain the action for the fluctuation fields by shifting the original fields by 
their expectation values~(\ref{eq:AE}) as 
\begin{equation}
\phi_n(k)\to \overline{\phi_n(k)}+\phi_n(k),\quad u(k)\to\overline{u(k)}+u(k), \quad 
\cdots \label{eq:AF}
\end{equation}
with a slight abuse of notation. Now, we do not impose the constraints $\phi_n=\phi_{-n}$ on 
$\phi_n$'s because $\phi_n$ and $\phi_{-n}$ are distinct fields carrying the opposite Kaluza-Klein 
charge to each other. We discuss here the Lorentz decompositions of the vector and the tensor 
fluctuation fields, according to~\cite{KS}. The massive vector field $\check{B}_{\mu}(k)$ 
whose polarization is tangential to the lump is divided into the longitudinal and the 
transverse parts as 
\begin{eqnarray}
\check{B}_{\mu}(k)&=&\check{B}^L_{\mu}(k)+\check{B}_{\mu}^T(k), \nonumber \\
\check{B}_{\mu}^L(k)&=&\frac{k_{\mu}}{ik^2}ik^{\nu}\check{B}_{\nu}(k). \label{eq:AG}
\end{eqnarray}
It is easily found that 
\[ k^{\mu}\check{B}_{\mu}(k)=k^{\mu}\check{B}_{\mu}^L(k), \quad k^{\mu}\check{B}_{\mu}^T(k)=0. \]
We further define  
\begin{equation}
\cB^L(k)\equiv i\sqrt{\ap}k^{\nu}\check{B}_{\nu}(k), \label{eq:AH}
\end{equation}
which can be regarded as a scalar in addition to the $X$-th vector component $\check{B}_X(k)$ 
transverse to the lump, and satisfies the reality condition
\[ \cB^L(k)^*=-i\sqrt{\ap}k^{\nu}\check{B}_{\nu}(k)^*=i\sqrt{\ap}(-k)^{\nu}\check{B}_{\nu}(-k)
=\cB^L(-k). \]
Using this definition, it follows that 
\begin{equation}
\check{B}_{\mu}(k)\check{B}^{\mu}(-k)=\check{B}_{\mu}^T(k)\check{B}^{T\mu}(-k)+\frac{1}{\ap k^2}
\cB^L(k)\cB^L(-k). \label{eq:AI}
\end{equation}
Note that the $\cB^L$-field has non-standard normalization, though it will not affect our 
analysis.\footnote{This problem could be avoided if we defined $\cB^L$ by $k^{\mu}\check{B}_{\mu}(k)
/\sqrt{|k^2|}$ instead of eq.(\ref{eq:AH}).}
In what follows, we will discard the first term in the right hand side of~(\ref{eq:AI}). 
Similarly, the tensor field $B_{MN}(k)$ is decomposed into several parts according to their 
Lorentz transformation properties as 
\begin{itemize}
	\item $XX$ component: $B_{XX}(k)$, 
	\item longitudinal component of the `Kaluza-Klein' vector $B_{\mu X}(k)$: 
	\begin{equation}
	B^L_{\mu X}(k)=\frac{k_{\mu}}{ik^2}\frac{\beta^L_X(k)}{\sqrt{\ap}}; \quad
	\beta^L_X(k)\equiv i\sqrt{\ap}k^{\nu}B_{\nu X}(k), \label{eq:AJ}
	\end{equation}
	\[B_{\mu X}(k)B^{\mu X}(-k)\longrightarrow \frac{1}{\ap k^2}\beta^L_X(k)\beta^L_X(-k), \]
	\item trace part of $B_{\mu\nu}(k)$: $\eta^{\mu\nu}_{(25)}B_{\mu\nu}(k)\equiv\sqrt{25}B(k)$, 
	\begin{equation}
	B_{\mu\nu}(k)=b_{\mu\nu}(k)+\frac{1}{5}\eta^{(25)}_{\mu\nu}B(k), \label{eq:AK}
	\end{equation}
	($b_{\mu\nu}$ is the traceless part of $B_{\mu\nu}$),
	\item longitudinal part of $b_{\mu\nu}(k)$: 
	\begin{equation}
	b^{LL}(k)\equiv\frac{k^{\rho}k^{\sigma}}{k^2}b_{\rho\sigma}(k),\quad 
	b^{LL}_{\mu\nu}(k)=\frac{k_{\mu}k_{\nu}}{k^2}b^{LL}(k), \label{eq:AL}
	\end{equation}
	\item and two transverse vectors and a symmetric 2-tensor.
\end{itemize}
Then we have 
\begin{eqnarray*}
B_{\mu\nu}(k)B^{\mu\nu}(-k)&\mathop{\longrightarrow}\limits^{\mathrm{scalar}}&B(k)B(-k)+
b^{LL}(k)b^{LL}(-k), \\ k^{\mu}k^{\nu}B_{\mu\nu}(k)&=&k^2b^{LL}(k)+\frac{1}{5}k^2 B(k). 
\end{eqnarray*}
Collecting the pieces above, there are 12 scalar fields we have to consider up to level 2 from 
the point of view of the lump world-volume, namely 
\begin{equation}
V^T=\biggl(\phi_2,\ \phi_1,\ \phi_0,\ \phi_{-1},\ \phi_{-2},\ u,\ \check{B}_X,\ \cB^L,\ B_{XX},\ 
\beta^L_X,\ B,\ b^{LL}\biggr) \label{eq:AM}
\end{equation}
in the vector notation with $T$ denoting the transposition. 

\begin{figure}[htbp]
	\begin{center}
	\includegraphics{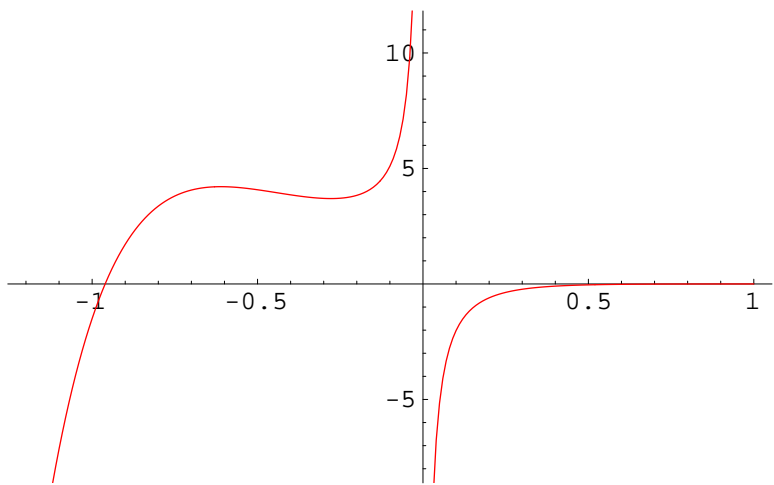}
	\end{center}
	\caption{The plot of $\det M$ as a function of $\mu=-\ap k^2\in [-1.2\ ,\ 1]$.}
	\label{fig:FK}
\end{figure}

\begin{figure}[htbp]
	\begin{center}
	\includegraphics{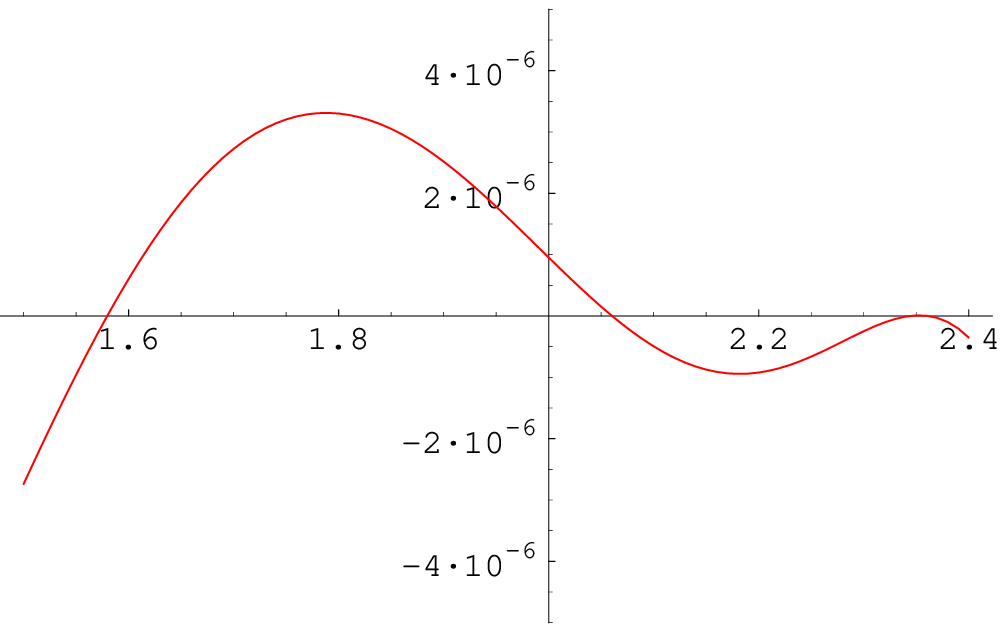}
	\end{center}
	\caption{The plot of $\det M$ as a function of $\mu=-\ap k^2\in [1.5\ ,\ 2.4]$.}
	\label{fig:FL}
\end{figure}

\begin{figure}[htbp]
	\begin{center}
	\includegraphics{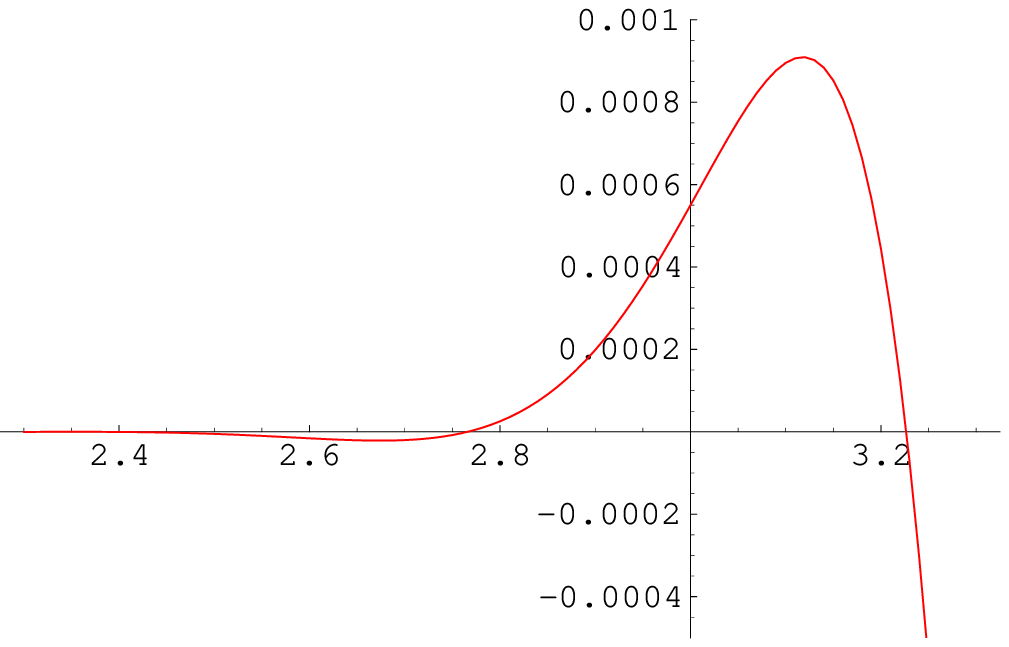}
	\end{center}
	\caption{The plot of $\det M$ as a function of $\mu=-\ap k^2\in [2.3\ ,\ 3.3]$.}
	\label{fig:FM}
\end{figure}

\begin{figure}[htbp]
	\begin{center}
	\includegraphics{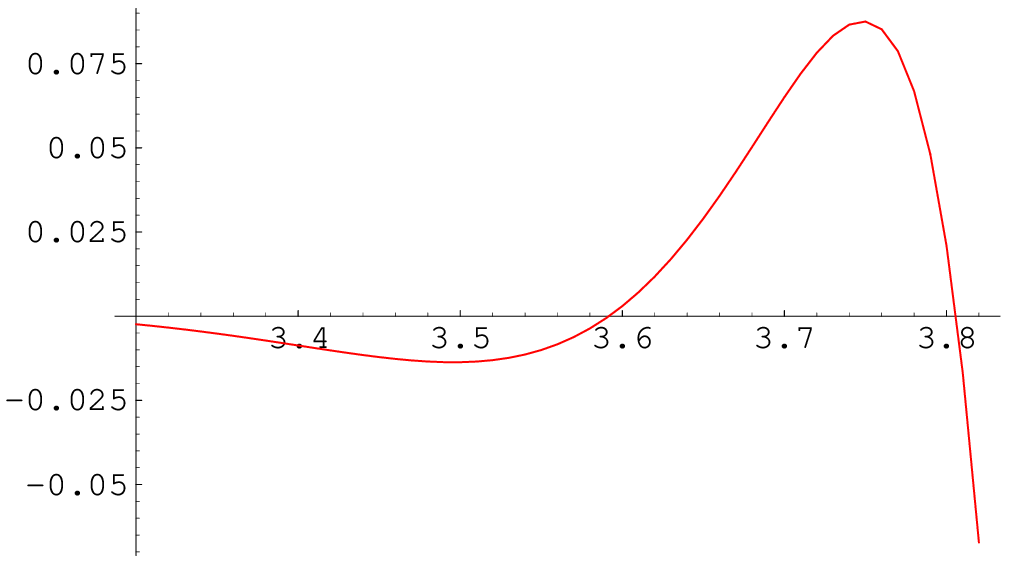}
	\end{center}
	\caption{The plot of $\det M$ as a function of $\mu=-\ap k^2\in [3.3\ ,\ 3.82]$.}
	\label{fig:FN}
\end{figure}

\begin{figure}[htbp]
	\begin{center}
	\includegraphics{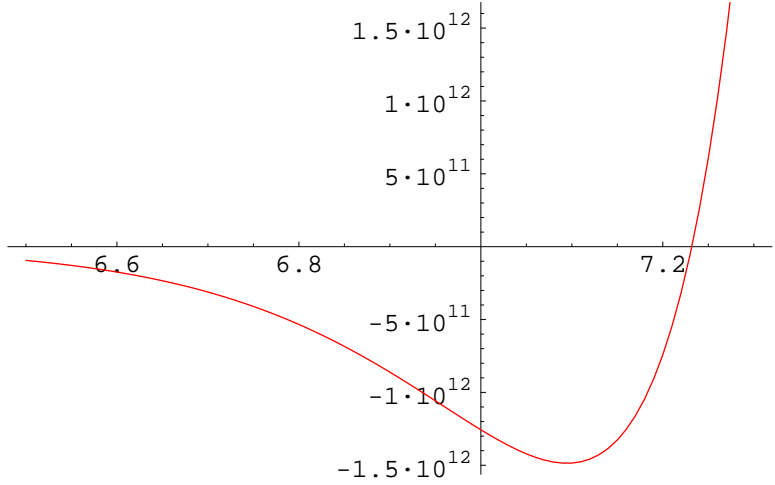}
	\end{center}
	\caption{The plot of $\det M$ as a function of $\mu=-\ap k^2\in [6.5\ ,\ 7.3]$.}
	\label{fig:FO}
\end{figure}

In the action for the fluctuation fields expanded around the lump solution, there are quadratic 
terms of the form\footnote{Terms of the form (vev$)^3$ only contribute to the tension of the lump, 
while (vev$)^2\cdot$(field) terms vanish due to the equations of motion.} 
\[ (\mathrm{vev})\cdot (\mathrm{field})\cdot (\mathrm{field}^{\prime}) \]
arising from the cubic interaction terms in the original action. Since the whole set of the cubic 
interaction terms includes almost all of the possible couplings, the quadratic form for the 
fluctuation fields is not diagonalized at all. Using the notation introduced above, the 
quadratic form can generally be written as 
\begin{equation}
(2\pi)^{26}R\int d^{25}k\ V^{\dagger}(-k)M(k^2)V(k) \label{eq:AN}
\end{equation}
where the `quadratic form matrix' $M$ is Hermitian and 
\[ V^{\dagger}=\biggl(\phi_{-2},\ \phi_{-1},\ \phi_0,\ \phi_1,\ \phi_2,\ u,\ \cdots\biggr). \]
Hence we can determine the scalar fluctuation spectrum on the lump by finding the set of values of 
$m^2=-k^2$ at which one (or some) of the eigenvalues of the $12\times 12$ matrix $M(k^2)$ vanishes. 
The explicit expressions for the components $M_{ij}$ of the matrix $M$ are displayed in 
Appendix~\ref{sec:appC}. Instead of diagonalizing this huge matrix, we have calculated the 
determinant of $M$ and looked for the values of $m^2=-k^2$ at which $\det M$ vanishes. 
We have plotted $\det M$ as a function of $\mu\equiv -\ap k^2$ in Figure~\ref{fig:FK}--\ref{fig:FO}: 
We had to divide the whole curve into several sectors because the scale of $\det M$ greatly 
changes from place to place. 
The seeming divergence of $\det M$ at $k^2=0$ (Figure~\ref{fig:FK}) is an artifact caused by the 
non-standard normalizations of $\cB^L$ and $\beta^L_X$. 
The numerical values of $\mu=\ap m^2=-\ap k^2$ where the $\det M$ vanishes have turned out to be 
\begin{eqnarray*}
& &\left.
	\begin{array}{|c|c|c|c|c|c|}
	\hline
	\mathrm{1st} & \mathrm{2nd} & \mathrm{3rd} & \mathrm{4th} & \mathrm{5th} & \mathrm{6th} 
	\\ \hline -0.960696 & 1.57958 & 2.06040 & 2.34365 & 2.36161 & 2.76601 \\ \hline
	\end{array}
\right. \\ & &\left.
	\begin{array}{|c|c|c|c|c|}
	\hline
	\mathrm{7th} & \mathrm{8th} & \mathrm{9th} & \mathrm{10th} & \mathrm{11th} \\ \hline 
	\ 3.22656 \quad & 3.59172 & 3.80594 & 7.23182 & \ \ 22.1 \quad \\ \hline
	\end{array}
\right. .
\end{eqnarray*}
The first point we should note is that there exists only one tachyonic state whose mass 
squared $\ap m^2\simeq -0.961$ is very close to the expected value 
$(\ap m^2=-1)$\textit{!} One may 
take it for granted that the properties of the original tachyon persist to the lump, but it is not 
correct because the tachyon field living on the lump world-volume is not identical to 
the original tachyon $\phi$ on the D25-brane at all. In fact, we must take suitable linear 
combinations of the fluctuation fields to diagonalize the matrix $M(k^2)$. Explicitly, the lump 
tachyon $\varphi$ takes the form 
\begin{eqnarray}
\hspace{-5mm}
\varphi&=&0.119537\ \phi_2+0.478126\ \phi_1+0.715829\ \phi_0+0.478126\ \phi_{-1}+0.119537\ \phi_{-2} 
\label{eq:AO} \\& &{}+0.00234923\ u-0.00332686\ \cB^L+0.0398185\ B_{XX}-0.0127727\ B+0.00604012\ b^{LL}. 
\nonumber
\end{eqnarray}
As we increase the level of approximation, more and more fields take part in constructing 
the lump tachyon, though their contribution will be small. Hence, our result that there 
is a tachyon with $\ap m^2\simeq -0.96$ on the lump world-volume is not only desirable 
but also quite non-trivial. 

The second point is that there are also several zeroes around $\ap m^2=2$.\footnote{Excessively 
heavier ones (the 10th and the 11th) than the truncation scale ($\ap m^2=1$) are not reliable.} 
Although some of these states may be $Q$-exact and hence not physical, we have not pursued 
this issue because the $Q$-exactness here is less significant than in the case of 
the tachyon vacuum. The spectrum on a D24-brane 
contains degenerate scalar states at $\ap m^2=1$ as in eq.(\ref{eq:AM}), 
so that the appearance of the nearly degenerate states on the lump seems to be 
consistent with the expected spectrum, though the values of $\ap m^2$ are a little too large: 
This point will be further discussed in section~\ref{sec:disc}. Actually, even the fact that 
the $\det M$, which is a very complicated function of $\mu=-\ap k^2$, has so many zeroes 
on the positive real axis is rather surprising and should not be an accident. 
Therefore, our result can be regarded as 
a strong piece of evidence that the fluctuation spectrum around the tachyonic lump solution 
agrees with that of a D24-brane. 

\section{Conclusions and Discussions}\label{sec:disc}
In this paper, we have discussed two issues concerned with the tachyonic lump solutions in bosonic cubic 
string field theory. One is the problem of whether other gauges than the Feynman-Siegel gauge may 
work as well and, if so, how the properties (tension, size) of the lump 
depend on these gauge choices. 
We have obtained the result that at level (2,4) we have succeeded in constructing lump solutions 
in various gauges except for the $B$-gauge, and that not only the tension but also the size 
of the lump is independent of these good gauge choices. Besides, solving the equations of motion 
without gauge fixing does not seem to give sensible results. 
It remains, however, to be resolved whether 
these conclusions, together with the legitimacy of the gauge fixing conditions~(\ref{eq:M}) 
themselves, persist to the higher levels or not. The other is to find the fluctuation spectrum 
around the tachyonic lump. Our calculations have shown that there are a tachyon with $\ap m^2
\simeq -0.96$ and some massive scalar states on the lump, and we regard this result as one more 
piece of evidence that a tachyonic lump solution represents a lower-dimensional D-brane. 
We conclude this paper with some discussions. 
\medskip

In the analysis of the fluctuation spectrum, we have considered the determinant of $M$. When we 
increase the truncation level, however, $M$ constitutes only part of the entire quadratic form matrix 
\[ \widetilde{M}=\left(
	\begin{array}{cc}
	M & L \\ L^{\dagger} & N
	\end{array}
\right), \]
and the masses of the states will be shifted due to the existence of non-vanishing off-diagonal block $L$, 
as well as the small corrections to $M$ itself. 
Nevertheless we expect that the qualitative features of our result will not be altered at higher levels 
because the contribution from $L$ is considered to be small. Indeed, since each component of $L$ comes from 
the cubic coupling among light fields and heavy fields, it is plausible that the 
coefficient of such a coupling is small. But recall that the values of $\ap m^2 (>0)$ 
at which $\det M$ vanishes were a little 
larger than the expected values. 
We hope that the masses of these massive states approach 1 as we increase 
the truncation level, whereas the mass of the tachyon scarcely changes. We cannot prove this 
statement because to reach the next level $\left(\frac{7}{3},\frac{14}{3}\right)$ 
requires us to evaluate the $26\times 26$ quadratic form matrix and its determinant, 
which has not been done yet. 
\smallskip

In the case of superstring, a tachyonic kink solution on a non-BPS D-brane was constructed 
in~\cite{KO2} by applying the level truncation scheme to superstring field theory formulated 
by Berkovits. Since the kink solution is to be identified with a BPS D-brane of one lower 
dimension, the fluctuation spectrum around it is expected to contain a massless scalar field 
representing the translational mode of the kink, instead of a tachyon. Although the actual 
calculations must become much more complicated because of the interaction terms of higher orders, 
it would in principle be possible to repeat similar calculations to those explained in this paper, 
and we believe that our expectations could be verified by explicit calculations. 

\section*{Acknowledgements}
I am grateful to Teruhiko Kawano for careful reading of the manuscript and many instructive comments. 
I would also like to thank Tohru Eguchi, Tadashi Takayanagi and Kazuhiro Sakai 
for helpful discussions. 

\newpage

\section*{Appendices}
\renewcommand{\thesection}{\Alph{section}}
\setcounter{section}{0}

\section{Expectation Values and Fitting of the Lumps}\label{sec:appA}
\begin{table}[htbp]
	\begin{center}
	\begin{tabular}{|c||c|c|c||c|c|c|}
	\hline
	\multicolumn{1}{|c||}{} & \multicolumn{3}{c||}{Gauge unfixed} & \multicolumn{3}{c|}
	{Feynman-Siegel gauge} \\ \cline{2-7}
	Field & vacuum & $R=\sqrt{3\ap}$ & $R=\sqrt{8\ap}$ & vacuum & 
	$R=\sqrt{3\ap}$ & $R=\sqrt{8\ap}$ \\ \hline \hline
	$\tau_0$ & 0.570140 & 0.261307 & 0.389790 & 0.541591 & 0.257030 & 0.366958 \\ \hline
	$\tau_1$ & 0 & $-0.389854$ & $-0.333837$ & 0 & $-0.384575$ & $-0.295236$ \\ \hline
	$\tau_2$ & 0 & $-0.109253$ & $-0.210478$ & 0 & $-0.107424$ & $-0.199225$ \\ \hline
	$\tau_3$ & 0 & --- & $-0.100522$ & 0 & --- & $-0.0877869$ \\ \hline
	$\tau_4$ & 0 & --- & $-0.0334873$ & 0 & --- & $-0.0281938$ \\ \hline
	$u$ & $-0.0878149$ & 0.286261 & 0.577919 & 0.173264 & 0.0888087 & 0.122846 \\ \hline
	$v$ & $-0.0390313$ & 0.0541575 & 0.161105 & 0.0518987 & $-0.00675676$ & 0.0195679 \\ \hline
	$w$ & $-0.0390313$ & 0.0935772 & 0.185541 & 0.0518987 & 0.0317837 & 0.0407204 \\ \hline
	$r$ & 0.182205 & $-0.0887169$ & $-0.230081$ & 0 & 0 & 0 \\ \hline
	\end{tabular}
	\end{center}
	\caption{The expectation values for the closed string vacuum solution and the 
	lump solutions for $R=\sqrt{3\ap}$ and for $R=\sqrt{8\ap}$ in the case of free gauge 
	and of the Feynman-Siegel gauge.}
	\label{tab:TAa}
\end{table}
\begin{table}[htbp]
	\begin{center}
	\begin{tabular}{|c||c|c|c||c|c|c|}
	\hline
	\multicolumn{1}{|c||}{} & \multicolumn{3}{c||}{$b_1|\Phi\rangle=0$} & \multicolumn{3}{c|}
	{$L_2^X|\Phi\rangle=0$} \\ \cline{2-7}
	Field & vacuum & $R=\sqrt{3\ap}$ & $R=\sqrt{8\ap}$ & vacuum & 
	$R=\sqrt{3\ap}$ & $R=\sqrt{8\ap}$ \\ \hline \hline
	$\tau_0$ & 0.547091 & 0.245390 & 0.354999 & 0.531880 & 0.261384 & 0.367602 \\ \hline
	$\tau_1$ & 0 & $-0.372937$ & $-0.286657$ & 0 & $-0.383699$ & $-0.289441$ \\ \hline
	$\tau_2$ & 0 & $-0.103975$ & $-0.200172$ & 0 & $-0.105811$ & $-0.197050$ \\ \hline
	$\tau_3$ & 0 & --- & $-0.0881422$ & 0 & --- & $-0.0847774$ \\ \hline
	$\tau_4$ & 0 & --- & $-0.0282572$ & 0 & --- & $-0.0270753$ \\ \hline
	$u$ & 0 & 0 & 0 & 0.228984 & 0.197288 & 0.216846 \\ \hline
	$v$ & $-0.00986666$ & $-0.0383939$ & $-0.0256342$ & 0 & 0 & 0 \\ \hline
	$w$ & $-0.00986666$ & $-0.00167938$ & $-0.00459855$ & 0.0709321 & 
	0.0657133 & 0.0705697 \\ \hline
	$r$ & 0.118928 & 0.0464671 & 0.0716666 & $-0.0356672$ & $-0.0484534$ & $-0.0476229$ 
	\\ \hline
	\end{tabular}
	\end{center}
	\caption{The expectation values for the closed string vacuum solution and the 
	lump solutions for $R=\sqrt{3\ap}$ and for $R=\sqrt{8\ap}$ in the case of $b_1$-gauge 
	and of $L_2^X$-gauge.}
	\label{tab:TAb}
\end{table}
\begin{table}[htbp]
	\begin{center}
	\begin{tabular}{|c||c|c|c||c|c|c|}
	\hline
	\multicolumn{1}{|c||}{} & \multicolumn{3}{c||}{$L_2^{\cM}|\Phi\rangle=0$} & \multicolumn{3}{c|}
	{$B|\Phi\rangle=0$} \\ \cline{2-7}
	Field & vacuum & $R=\sqrt{3\ap}$ & $R=\sqrt{8\ap}$ & vacuum & 
	$R=\sqrt{3\ap}$ & $R=\sqrt{8\ap}$ \\ \hline \hline
	$\tau_0$ & 0.544185 & 0.245507 & 0.356298 & 0.632548 & 0.237673 & 0.363251 \\ \hline
	$\tau_1$ & 0 & $-0.374302$ & $-0.288654$ & 0 & $-0.335882$ & $-0.283967$ \\ \hline
	$\tau_2$ & 0 & $-0.104706$ & $-0.200508$ & 0 & $-0.0860061$ & $-0.210525$ \\ \hline
	$\tau_3$ & 0 & --- & $-0.0886644$ & 0 & --- & $-0.0899609$ \\ \hline
	$\tau_4$ & 0 & --- & $-0.0284618$ & 0 & --- & $-0.0295582$ \\ \hline
	$u$ & 0.0249170 & $-0.00187157$ & 0.00517107 & $-0.214818$ & $-0.195351$ & $-0.235178$ 
	\\ \hline $v$ & 0.00155036 & $-0.0334123$ & $-0.0174624$ & $-0.0919081$ & 
	$-0.0874468$ & $-0.0948774$ \\ \hline
	$w$ & 0 & 0 & 0 & $-0.0919081$ & $-0.0574067$ & $-0.0727994$ \\ \hline
	$r$ & 0.0965187 & 0.0392659 & 0.0589039 & 0.316274 & 0.118837 & 0.181625 \\ \hline
	\end{tabular}
	\end{center}
	\caption{The expectation values for the closed string vacuum solution and the 
	lump solutions for $R=\sqrt{3\ap}$ and for $R=\sqrt{8\ap}$ in the case of $L_2^{\cM}$-gauge 
	and of $B$-gauge.}
	\label{tab:TAc}
\end{table}
\begin{table}[htbp]
	\begin{center}
	\begin{tabular}{|c||c|c|c||c|c|c|}
	\hline
	\multicolumn{1}{|c||}{} & \multicolumn{3}{c||}{$R=\sqrt{3\ap}$} & \multicolumn{3}{c|}
	{$R=2\sqrt{2\ap}$} \\ \cline{2-7}
	Gauge & $a$ & $b$ & $\sigma$ & $a$ & $b$ & $\sigma$ \\ \hline \hline
	unfixed & $-0.562279$ & 0.809903 & 1.61433 & $-0.584817$ & 0.881559 & 1.56733 \\ \hline
	Feynman-Siegel & $-0.554191$ & 0.798791 & 1.61608 & $-0.542040$ & 0.796381 & 1.55317 \\ \hline
	$u=0$ & $-0.533709$ & 0.774535 & 1.61711 & $-0.525443$ & 0.786111 & 1.53030 \\ \hline
	$v=0$ & $-0.558916$ & 0.796413 & 1.62301 & $-0.539906$ & 0.781668 & 1.55641 \\ \hline
	$w=0$ & $-0.534616$ & 0.777515 & 1.61531 & $-0.527755$ & 0.789999 & 1.53214 \\ \hline
	$B=0$ & $-0.503540$ & 0.694859 & 1.66272 & $-0.533721$ & 0.799624 & 1.50111 \\ \hline
	\end{tabular}
	\end{center}
	\caption{The results of fitting $-t(x)$ in eq.(\ref{eq:U}) with $G(x)$ 
	in eq.(\ref{eq:V}).}
	\label{tab:TAd}
\end{table}

\section{Level (2,4)-truncated Action}\label{sec:appB}
We write below the explicit expression of the action obtained by substituting the string 
field~(\ref{eq:AC}) into the cubic action~(\ref{eq:G}). 
\begin{eqnarray*}
\hspace{-8mm}
\frac{-g_o^2S}{(2\pi)^{26}R}&=&\int d^{25}\!k\Biggl[\frac{1}{2}\sum_{n=-2}^2\left(\frac{\ap}{R^2}
n^2+\ap k^2-1\right)\phi_n(k)\phi_{-n}(-k)-\frac{1}{2}(\ap k^2+1)u(k)u(-k) \\
& &+(\ap k^2+1)B_{\mu\nu}(k)B^{\mu\nu}(-k)+2(\ap k^2+1)B_{\mu X}(k)B^{\mu X}(-k) \\
& &+(\ap k^2+1)B_{XX}(k)B_{XX}(-k) \\
& &+(\ap k^2+1) \check{B}_{\mu}(k)\check{B}^{\mu}(-k)+(\ap k^2+1)\check{B}_X(k)\check{B}_X
(-k)\Biggr] \\ & &+\int d^{25}\!k_1d^{25}\!k_2d^{25}\!k_3\ \delta^{25}\left(\sum_{i=1}^3k_i\right)
K^3\Biggl[\frac{1}{3}
\sum_{n_1,n_2,n_3=-2}^2\delta_{n_1+n_2+n_3}\widetilde{\phi}_{n_1}(k_1)
\widetilde{\phi}_{n_2}(k_2)\widetilde{\phi}_{n_3}(k_3) \\
& &+\frac{11}{27}\sum_{n=-1}^1\widetilde{\phi}_n(k_1)\widetilde{\phi}_{-n}(k_2)\widetilde{u}(k_3)
-\frac{2}{9}\sqrt{2\ap}\sum_{n=-1}^1\widetilde{\phi}_n(k_1)ik_2^{\mu}\widetilde{\check{B}}_{\mu}
(k_2)\widetilde{\phi}_{-n}(k_3) \\ 
& &+\frac{1}{27}\sum_{n=-1}^1\widetilde{\phi}_n(k_1)\widetilde{\phi}_{-n}(k_3)\Biggl\{ 8\ap 
(k_1-k_3)^{\mu}(k_1-k_3)^{\nu}\widetilde{B}_{\mu\nu}(k_2)+\frac{32\ap}{R}n(k_1-k_3)^{\mu}
\widetilde{B}_{\mu X}(k_2) \\
& &+\frac{32\ap}{R^2}n^2\widetilde{B}_{XX}(k_2)-5\widetilde{B}_{XX}(k_2)-5 \eta^{\mu\nu}_{(25)}
\widetilde{B}_{\mu\nu}(k_2)\Biggr\} \\ 
& &+\frac{19}{243}\widetilde{\phi}_0(k_1)\widetilde{u}(k_2)\widetilde{u}(k_3)+\frac{8\ap}{81}
\widetilde{\phi}_0(k_1)ik_2^{\mu}\widetilde{\check{B}}_{\mu}(k_2)ik_3^{\nu}
\widetilde{\check{B}}_{\nu}(k_3) \\
& &-\frac{256}{243}\widetilde{\phi}_0(k_1)\left(i\widetilde{\check{B}}_X(k_2)i
\widetilde{\check{B}}_X(k_3)+\eta^{\mu\nu}_{(25)}i\widetilde{\check{B}}_{\mu}(k_2)i
\widetilde{\check{B}}_{\nu}(k_3)\right) \\
& &+\frac{1}{729}\widetilde{\phi}_0(k_1)\left\{ 8\ap (k_2-k_1)^{\rho}(k_2-k_1)^{\sigma}
\widetilde{B}_{\rho\sigma}(k_3)-5\widetilde{B}_{XX}(k_3)-5\eta^{\rho\sigma}_{(25)}\widetilde{B}_{\rho
\sigma}(k_3)\right\} \\ & &\quad \times\left\{ 8\ap (k_1-k_3)^{\mu}(k_1-k_3)^{\nu}
\widetilde{B}_{\mu\nu}(k_2)-5\widetilde{B}_{XX}(k_2)-5\eta^{\mu\nu}_{(25)}\widetilde{B}_{\mu
\nu}(k_2)\right\} \\
& &+\frac{512}{729}\widetilde{\phi}_0(k_1)\left\{ \widetilde{B}_{XX}(k_2)\widetilde{B}_{XX}(k_3)
+\widetilde{B}_{\mu\nu}(k_2)\widetilde{B}^{\mu\nu}(k_3)+2\eta^{\mu\rho}_{(25)}
\widetilde{B}_{\mu X}(k_2)\widetilde{B}_{\rho X}(k_3) \right\} \\
& &+\frac{512}{729}\ap \widetilde{\phi}_0(k_1)(k_1-k_3)^{\mu}(k_2-k_1)^{\sigma}\left\{
\widetilde{B}_{\mu X}(k_2)\widetilde{B}_{\sigma X}(k_3)+\eta^{\nu\rho}_{(25)}
\widetilde{B}_{\mu\nu}(k_2)\widetilde{B}_{\rho\sigma}(k_3)\right\} \\ 
& &+\frac{22}{729}\left\{ 8\ap (k_2-k_1)^{\mu}(k_2-k_1)^{\nu}\widetilde{B}_{\mu\nu}(k_3)
-5\widetilde{B}_{XX}(k_3)-5\eta^{\mu\nu}_{(25)}\widetilde{B}_{\mu\nu}(k_3)\right\}
\widetilde{\phi}_0(k_1)\widetilde{u}(k_2) \\
& &-\frac{44\sqrt{2\ap}}{243}\widetilde{\phi}_0(k_1)\widetilde{u}(k_2)ik_3^{\mu}
\widetilde{\check{B}}_{\mu}(k_3)+\frac{512}{729}\sqrt{2\ap}\widetilde{\phi}_0(k_1)\biggl\{
(k_2-k_1)^{\rho}i\widetilde{\check{B}}_X(k_2)\widetilde{B}_{\rho X}(k_3) \\ 
& &\quad +(k_2-k_1)^{\rho}\eta^{\mu\nu}i\widetilde{\check{B}}_{\mu}(k_2)\widetilde{B}_{\nu\rho}
(k_3)\biggr\} -\frac{4\sqrt{2\ap}}{243}\widetilde{\phi}_0(k_1)ik_2^{\mu}
\widetilde{\check{B}}_{\mu}(k_2) \\ & &\quad \times
\biggl\{ 8\ap (k_2-k_1)^{\nu}(k_2-k_1)^{\rho}\widetilde{B}_{\nu\rho}(k_3)-5\widetilde{B}_{XX}
(k_3)-5\eta^{\nu\rho}_{(25)}\widetilde{B}_{\nu\rho}(k_3)\biggr\}\Biggr],
\end{eqnarray*}
where we have defined the tilded fields as 
\begin{eqnarray*}
\widetilde{\phi}_n(k)&=&K^{-\frac{\ap}{R^2}n^2-\ap k^2}\phi_n(k), \\
\widetilde{u}(k)&=&K^{-\ap k^2}u(k),
\end{eqnarray*}
and so on.

\section{Quadratic Form Matrix $M(k^2)$}\label{sec:appC}
We show the explicit form of the matrix $M$ defined in eq.(\ref{eq:AN}). 
The assignment of indices follows from $V$ in eq.(\ref{eq:AM}). 
For simplicity, we omit the overlines on the understanding that all the field variables 
appearing below represent their expectation values. 
\begin{eqnarray*}
M_{11}&=&\frac{1}{2}\left(\ap k^2+\frac{4\ap}{R^2}-1\right)+\phi_0 K^{3-2\ap k^2-\frac{8\ap}{R^2}}, \\
M_{22}&=&\frac{1}{2}\left(\ap k^2+\frac{\ap}{R^2}-1\right)+\phi_0 K^{3-2\ap k^2-\frac{2\ap}{R^2}}+
\frac{11}{27}u K^{3-2\ap k^2-\frac{2\ap}{R^2}} \\ & &+\frac{1}{27}\left[\left(16\ap k^2-\frac{125}{2}
\right)w+\left(\frac{16\ap}{R^2}-\frac{5}{2}\right)v\right]K^{3-2\ap k^2-\frac{2\ap}{R^2}}, \\
M_{33}&=&\frac{1}{2}\left(\ap k^2-1\right)+\phi_0 K^{3-2\ap k^2}+
\frac{11}{27}u K^{3-2\ap k^2} \\ & &+\frac{1}{27}\left[\left(16\ap k^2-\frac{125}{2}
\right)w-\frac{5}{2}v\right]K^{3-2\ap k^2}, \\
M_{44}&=&\frac{1}{2}\left(\ap k^2+\frac{\ap}{R^2}-1\right)+\phi_0 K^{3-2\ap k^2-\frac{2\ap}{R^2}}+
\frac{11}{27}u K^{3-2\ap k^2-\frac{2\ap}{R^2}} \\ & &+\frac{1}{27}\left[\left(16\ap k^2-\frac{125}{2}
\right)w+\left(\frac{16\ap}{R^2}-\frac{5}{2}\right)v\right]K^{3-2\ap k^2-\frac{2\ap}{R^2}}, \\
M_{55}&=&\frac{1}{2}\left(\ap k^2+\frac{4\ap}{R^2}-1\right)+\phi_0 K^{3-2\ap k^2-\frac{8\ap}{R^2}}, \\
M_{12}&=&\phi_1K^{3-2\ap k^2-\frac{6\ap}{R^2}}, \\
M_{13}&=&\phi_2K^{3-2\ap k^2-\frac{8\ap}{R^2}}, \\
M_{14}&=&M_{15}=M_{25}=M_{41}=M_{51}=M_{52}=0, \\
M_{21}&=&\phi_{-1}K^{3-2\ap k^2-\frac{6\ap}{R^2}}, \\
M_{23}&=&\phi_1K^{3-2\ap k^2-\frac{2\ap}{R^2}}, \\
M_{24}&=&\phi_2K^{3-2\ap k^2-\frac{6\ap}{R^2}}, \\
M_{31}&=&\phi_{-2}K^{3-2\ap k^2-\frac{8\ap}{R^2}}, \\
M_{32}&=&\phi_{-1}K^{3-2\ap k^2-\frac{2\ap}{R^2}}, \\
M_{34}&=&\phi_1K^{3-2\ap k^2-\frac{2\ap}{R^2}}, \\
M_{35}&=&\phi_2K^{3-2\ap k^2-\frac{8\ap}{R^2}}, \\
M_{42}&=&\phi_{-2}K^{3-2\ap k^2-\frac{6\ap}{R^2}}, \\
M_{43}&=&\phi_{-1}K^{3-2\ap k^2-\frac{2\ap}{R^2}}, \\
M_{45}&=&\phi_1K^{3-2\ap k^2-\frac{6\ap}{R^2}}, \\
M_{53}&=&\phi_{-2}K^{3-2\ap k^2-\frac{8\ap}{R^2}}, \\
M_{54}&=&\phi_{-1}K^{3-2\ap k^2-\frac{6\ap}{R^2}}, \\
M_{66}&=&-\frac{1}{2}(\ap k^2+1)+\frac{19}{243}\phi_0K^{3-2\ap k^2}, \\
M_{67}&=&M_{76}=M_{6,10}=M_{10,6}=0, \\
M_{68}&=&M_{86}=-\frac{22\sqrt{2}}{243}\phi_0K^{3-2\ap k^2}, \\
M_{69}&=&M_{96}=-\frac{55}{729}\phi_0K^{3-2\ap k^2}, \\
M_{6,11}&=&M_{11,6}=\frac{11}{729}\phi_0K^{3-2\ap k^2}\left(\frac{8}{5}\ap k^2-25\right), \\
M_{6,12}&=&M_{12,6}=\frac{88}{729}\phi_0K^{3-2\ap k^2}\ap k^2, \\
M_{77}&=&\ap k^2+1+\frac{256}{243}\phi_0K^{3-2\ap k^2}, \\
M_{78}&=&M_{79}=M_{7,11}=M_{7,12}=M_{87}=M_{97}=M_{11,7}=M_{12,7}=0, \\
M_{7,10}&=&M_{10,7}=-\frac{256\sqrt{2}}{729}\phi_0K^{3-2\ap k^2}, \\
M_{88}&=&\frac{\ap k^2+1}{\ap k^2}+\left(\frac{8}{81}+\frac{256}{243\ap k^2}\right)\phi_0K^{3-2\ap k^2}, \\
M_{89}&=&M_{98}=\frac{10\sqrt{2}}{243}\phi_0K^{3-2\ap k^2}, \\
M_{8,10}&=&M_{10,8}=0, \\
M_{8,11}&=&M_{11,8}=\frac{256\sqrt{2}}{3645}\phi_0K^{3-2\ap k^2}-\frac{2\sqrt{2}}{243}\phi_0
K^{3-2\ap k^2}\left(\frac{8}{5}\ap k^2-25\right), \\
M_{8,12}&=&M_{12,8}=\frac{256\sqrt{2}}{729}\phi_0K^{3-2\ap k^2}-\frac{16\sqrt{2}}{243}\phi_0
K^{3-2\ap k^2}\ap k^2, \\
M_{99}&=&\ap k^2+1+\frac{537}{729}\phi_0K^{3-2\ap k^2}, \\
M_{9,10}&=&M_{10,9}=0, \\
M_{9,11}&=&M_{11,9}=\frac{1}{729}(-8\ap k^2+125)\phi_0K^{3-2\ap k^2}, \\
M_{9,12}&=&M_{12,9}=-\frac{40}{729}\ap k^2 \phi_0K^{3-2\ap k^2}, \\
M_{10,10}&=&\frac{2(\ap k^2+1)}{\ap k^2}+\frac{512}{729}\phi_0K^{3-2\ap k^2}+\frac{1024}{729\ap k^2}
\phi_0K^{3-2\ap k^2}, \\
M_{10,11}&=&M_{11,10}=M_{10,12}=M_{12,10}=0, \\
M_{11,11}&=&\ap k^2+1+\frac{1}{729}\phi_0K^{3-2\ap k^2}\left(-\frac{8}{5}\ap k^2+25\right)^2 \\
& &{}+\frac{512}{729}\phi_0K^{3-2\ap k^2}+\frac{512}{25\times 729}\phi_0K^{3-2\ap k^2}\ap k^2, \\
M_{11,12}&=&M_{12,11}=-\frac{8}{729}\ap k^2\left(-\frac{8}{5}\ap k^2+25\right)\phi_0K^{3-2\ap k^2}
+\frac{512}{3645}\phi_0K^{3-2\ap k^2}\ap k^2, \\
M_{12,12}&=&\ap k^2+1+\frac{1}{729}\phi_0K^{3-2\ap k^2}(64 \ap{}^2 (k^2)^2+512+512\ap k^2), \\
M_{16}&=&M_{17}=M_{18}=M_{19}=M_{1,10}=M_{1,11}=M_{1,12}=M_{61} \\
&=&M_{71}=M_{81}=M_{91}=M_{10,1}=M_{11,1}=M_{12,1}=0, \\
M_{56}&=&M_{57}=M_{58}=M_{59}=M_{5,10}=M_{5,11}=M_{5,12}=M_{65} \\
&=&M_{75}=M_{85}=M_{95}=M_{10,5}=M_{11,5}=M_{12,5}=0, \\
M_{26}&=&M_{62}=M_{64}=M_{46}=\frac{11}{27}\phi_{\pm 1}K^{3-2\ap k^2-\frac{2\ap}{R^2}}, \\
M_{27}&=&M_{72}=M_{47}=M_{74}=0, \\
M_{28}&=&M_{82}=M_{84}=M_{48}=-\frac{2\sqrt{2}}{9}\phi_{\pm 1}K^{3-2\ap k^2-\frac{2\ap}{R^2}}, \\
M_{29}&=&M_{92}=M_{94}=M_{49}=\frac{1}{27}\left(\frac{32\ap}{R^2}-5\right)\phi_{\pm 1}
K^{3-2\ap k^2-\frac{2\ap}{R^2}}, \\
M_{2,10}&=&-M_{10,2}=-M_{4,10}=M_{10,4}=
-\frac{32 \sqrt{\ap} i}{27 R}\phi_{\pm 1}K^{3-2\ap k^2-\frac{2\ap}{R^2}}, \\
M_{2,11}&=&M_{11,2}=M_{11,4}=M_{4,11}=\frac{1}{27}\left(\frac{8}{5}\ap k^2-25\right)\phi_{\pm 1}
K^{3-2\ap k^2-\frac{2\ap}{R^2}}, \\
M_{2,12}&=&M_{12,2}=M_{12,4}=M_{4,12}=\frac{8}{27}\ap k^2\phi_{\pm 1}K^{3-2\ap k^2-\frac{2\ap}{R^2}}, \\
M_{36}&=&M_{63}=\frac{11}{27}\phi_0K^{3-2\ap k^2}+\frac{19}{243}uK^{3-2\ap k^2}+\frac{11}{729}\left\{
\left(16\ap k^2-\frac{125}{2}\right)w-\frac{5}{2}v\right\} K^{3-2\ap k^2}, \\
M_{37}&=&M_{73}=0, \\
M_{38}&=&M_{83}=-\frac{2\sqrt{2}}{9}\phi_0K^{3-2\ap k^2}-\frac{22\sqrt{2}}{243}uK^{3-2\ap k^2}
+\frac{256\sqrt{2}}{729}w K^{3-2\ap k^2} \\ & &{}-\frac{2\sqrt{2}}{243}K^{3-2\ap k^2}
\left\{ \left( 16\ap k^2-\frac{125}{2}\right) w-\frac{5}{2}v\right\} , \\
M_{39}&=&M_{93}=-\frac{5}{27}\phi_0K^{3-2\ap k^2}-\frac{5}{729}K^{3-2\ap k^2}\left\{ \left(
16 \ap k^2-\frac{125}{2}\right) w-\frac{5}{2} v\right\} \\ & &{}+\frac{256}{729}vK^{3-2\ap k^2}
-\frac{55}{729}u K^{3-2\ap k^2}, \\
M_{3,10}&=&M_{10,3}=0, \\
M_{3,11}&=&M_{11,3}=\frac{1}{27}\left(\frac{8}{5}\ap k^2-25\right)\phi_0K^{3-2\ap k^2} \\
& &{}+\frac{1}{729}\left(\frac{8}{5}\ap k^2-25\right) K^{3-2\ap k^2}\left\{\left( 16\ap k^2
-\frac{125}{2}\right) w-\frac{5}{2}v\right\} \\ & &{}+\frac{256\times 5}{729}w K^{3-2\ap k^2}
-\frac{512}{5\times 729}wK^{3-2\ap k^2}\ap k^2+\frac{11}{729}u K^{3-2\ap k^2}\left(\frac{8}{5}\ap k^2
-25\right), \\
M_{3,12}&=&M_{12,3}=\frac{8}{27}\ap k^2\phi_0K^{3-2\ap k^2}+\frac{8}{729}\ap k^2 K^{3-2\ap k^2}
\left\{\left( 16\ap k^2-\frac{125}{2}\right) w-\frac{5}{2}v\right\} \\
& &{}-\frac{512}{729}wK^{3-2\ap k^2}\ap k^2+\frac{88}{729}u K^{3-2\ap k^2}\ap k^2.
\end{eqnarray*}


\end{document}